\documentclass[pre,aps,pacs,twocolumn]{revtex4}
\usepackage{epsfig}
\usepackage{graphicx}
\usepackage{bm}
\usepackage{color}
\newcommand{\B}[1]{{\bm{#1}}}

\usepackage[latin1]{inputenc}
\newcommand{\beq}{\begin{equation}}
\newcommand{\eeq}{\end{equation}}
\newcommand{\bea}{\begin{eqnarray}}
\newcommand{\eea}{\end{eqnarray}}

\begin{document}

\title{Theory of Relaxation Dynamics in Glass-Forming Hydrogen-Bonded Liquids }
\author{H.G.E.Hentschel$^{1,2}$ and Itamar Procaccia$^1$}
\affiliation{$^1$Dept. of Chemical Physics, The Weizmann Institute of
Science, Rehovot 76100, Israel
\\$^2$Dept of Physics,
Emory University, Atlanta, Georgia. }
\begin{abstract}
We address the relaxation dynamics in hydrogen-bonded super-cooled liquids near the glass transition, measured via Broad-Band Dielectric Spectroscopy (BDS). We propose a theory based on decomposing the relaxation of the macroscopic dipole moment into contributions from hydrogen bonded clusters of $s$ molecules, with 
$s_{min}\le s \le s_{max}$. The existence of $s_{max}$ is due to dynamical arrest and its value may depend on the cooling protocol and on the aging time. The existence of $s_{max}$ is translated into a sum-rule on the concentrations of clusters of size $s$. We construct the statistical mechanics of the super-cooled liquid subject to this sum-rule as a constraint, to estimate the temperature-dependent density of clusters of size $s$. With a theoretical estimate of the relaxation time of each cluster we provide predictions for the real and imaginary part of the frequency dependent dielectric response. The predicted spectra and their temperature dependence are in accord with measurements, explaining a host of phenomenological fits like the Vogel-Fulcher fit and the stretched exponential fit. Using glycerol as a particular example we demonstrate quantitative correspondence between theory and experiments. The theory also demonstrates that the $\alpha$ peak and the ``excess wing" stem from the same physics in this material. The theory also shows that in other hydrogen-bonded glass formers the ``excess wing" can develop into a $\beta$ peak, depending on the molecular material parameters (predominantly the surface energy of the clusters). We thus argue that $\alpha$ and $\beta$ peaks can stem from the same physics. Finally we address the BDS in constrained geometries (pores) and explain why recent experiments on glycerol did not show a deviation from bulk spectra.
\end{abstract}
\maketitle
\section{Introduction}

In this paper we propose a theory, based on a simple physical model at the mesoscale, that can account quantitatively for most of the observed features in the remarkable relaxation dynamics of glass-forming hydrogen-bonded liquids \cite{51DC,72JW,91Ang,91NRP,96LPDGBL,00SBLL,05PHRBFKB}. The paradigmatic example of such systems is dry glycerol and glycerol-water mixtures, but other alcohols and alcohol mixtures show similar properties. The relaxation dynamics of such systems in the
vicinity of the glass transition exhibit an extremely wide range of frequencies, spanning sometimes 16 orders of magnitude or more in the frequency domain. Despite the fact that good measurements have been available for more than half a century \cite{51DC}, and after considerable experimental and theoretical effort, including apparent ``first principles" theories like Mode-Coupling \cite{92GS}, there is no accepted derivation of the observed characteristics of the relaxation dynamics. Until now most discussions of experimental data are limited to fitting phenomenological expressions that are not the result of a proper theory \cite{06BFKHF}. The present paper attempts to close this gap.

An excellent technique to probe this enormous range of frequencies is broad-band dielectric spectroscopy (BDS) that covers the frequency band
($10^{-5}$-- $10^{11}$ Hz) and temperature range (100-600K). The interpretation of the measurements
in BDS is facilitated by the fact that the dielectric theory for BDS is well established and independent of the nature of the relaxational mechanisms involved in any particular material and at any particular temperature and pressure.  We can express the
 frequency dependent dielectric constant $\epsilon (\omega )$ in terms of the Laplace-Fourier transform 
 \begin{equation}
 \label{diel}
\frac{\epsilon (\omega ) - \epsilon_{\infty}}{ \epsilon_0 - \epsilon_{\infty}}= \int_0^{\infty}\left(\frac{-d\phi (t)}{dt} \right)e^{-i\omega t} + \frac{4\pi i \sigma_{dc}/(\epsilon_0 - \epsilon_{\infty} )}{\omega}.
 \end{equation}
 Here $\epsilon_{\infty}$ is the high frequency dielectric constant due to fast rotational processes above $10^{12}$Hz and
 atomic polarization; $\epsilon_0$ is the static dielectric constant; and $\sigma_{dc}$ is the dc conductivity of the medium. The response function $\phi (t)$ is given by the normalized  correlation function of the macroscopic dipole moment ${\bf M}(t)$ 
\begin{equation}
\label{response0}
\phi (t) = \frac{\langle {\bf M}(t)\cdot{\bf M}(0)\rangle}{\langle {\bf M}(0)\cdot{\bf M}(0)\rangle}.
\end{equation}

\begin{figure}
\centering
\epsfig{width=.5\textwidth,file=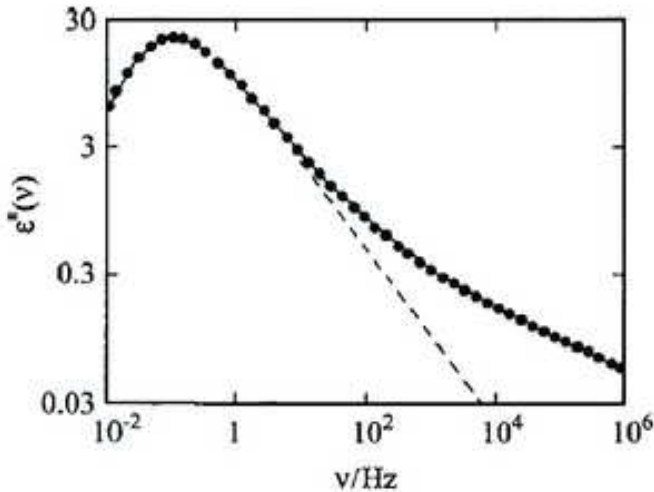}
\caption{A typical BDS absorption spectrum read from the imaginary part of $\epsilon(\omega)$ for dry glycerol at temperature $T=196$. The dashed line is the phenomenological Davidson-Cole formula, which fails to describe the so called ``excess wing".}
\label{BDSspec}
\end{figure}
\begin{figure}
\centering
\epsfig{width=.5\textwidth,file=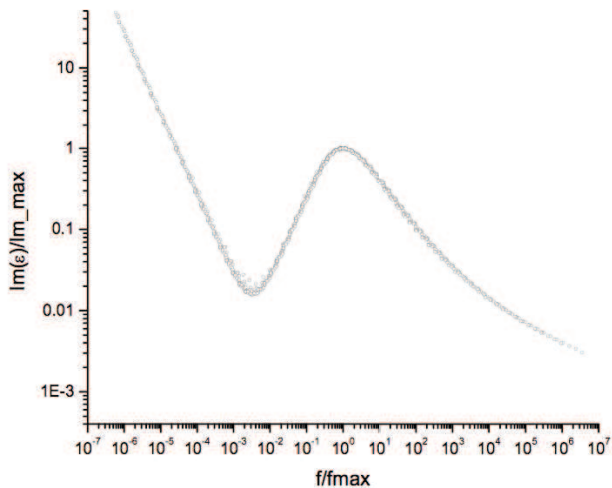}
\caption{Experimental BDS absorption spectra of dry glycerol at a range of temperatures 202-292 K, where frequencies were rescaled to the maximum of the $\alpha$ peak and amplitudes were rescaled to the amplitude ot he same peak. Note that in comparison to Fig. \ref{BDSspec} here the dc contribution is shown. The data collapse indicates strongly that the $\alpha$ peak,  the ``excess wing" stem and the low frequency conductivity contribution are due to  the same physical mechanism.} 
\label{rescaled}
\end{figure}

A typical BDS absorption spectrum read from the imaginary part of $\epsilon(\omega)$ with dry glycerol at temperature $T=196K$ is shown in Fig. \ref{BDSspec}.
One observes what has been termed in the literature "the main relaxation process" or $\alpha$ peak  at $f_{max}\approx 10^{-1}$Hz, and then tapering off initially with a typical power law form $\omega^{-\beta}$ which will appear as a straight line on a log-log plot, but is generally followed by a typical wing, which in Fig.~\ref{BDSspec}
has a gentle curvature at frequencies above $f \ge 10^2$Hz. This part of the spectrum had been termed in the literature "the Excess Wing"; This term does not reflect a deep realization of the existence of an excess loss, but rather the name stems simply from the inability of the phenomenological Davidson-Cole fit formula \cite{50DC} to agree with the experimental spectrum at this frequency range. While not in glycerol, in other hydrogen-bonded glass formers this wing can also exhibit a clear shoulder or even a distinct second $\beta$ peak \cite{00SBLL}. The literature does not agree on whether the $\alpha$ relaxation and the `excess wing' (or the $\beta$ peak when it exists) stem from the same physics or not \cite{00Nga}. A very strong experimental argument in favor of the very same physics was presented in \cite{05PHRBFKB} where spectra taken
at different temperatures were rescaled by the position and amplitude of the $\alpha$ peak, see Fig. \ref{rescaled}. The excellent data collapse is a strong indication that the $\alpha$ process and the ``excess wing" stem from the same physical mechanism. The theory presented below will corroborate this finding and will clarify the physical origin
of the $\alpha$ and $\beta$ peaks when both exist.

Of course, the raw spectra exhibit a strong temperature dependence. Using spectra measured in the range of temperatures $190-240K$ in dry glycerol, the positions of the maxima of the ``main process"  have been fitted to the Vogel-Fulcher formula by
defining $\tau_{max}=1/2\pi f_{max}$ and then writing
\begin{equation}
\label{VF}
\tau_{max}(T) =\tau_v \exp{[DT_v/(T-T_v)]} \ . 
\end{equation}
The three parameters were fitted to the data with the results $\ln \tau_v=-35.9$, $D=22.7$ and $T_v=125K$. Note that this fit implies an attempt time $\tau_v \approx 10^{-16}$ sec which
is a very short time indeed. We will argue below that the Vogel-Fulcher formula has no deep meaning, and the parameters involved are of limited physical interest. Although the spectra derived from our theory below will be shown to agree with Eq. (\ref{VF}) in the range of measurement used, we argue the formula is nothing but a data fit that can be used only in a finite range of temperatures. The prediction of the theory described below is that that there is no true divergence of $\tau_{max}$ at any temperature $T>0$.

The task of the theorist is then to derive, preferably from first principles, the form of the frequency dependent dielectric response $\epsilon(\omega)$ for this system, to explain the observed spectra (both real and imaginary parts). In particular a theory should indicate whether the ``main process" and the ``excess wing" stem from the same physics or does one need to invoke more than one relaxation mechanism. We will see that the former is the case here. Then one needs to explain the observed fits to the Vogel-Fulcher formula. While the present authors do not ascribe much physical significance to the parameters appearing in the formula, nevertheless the experimental fits should be accounted for, and hopefully the origin of the parameters identified. Finally one needs to explain the data collapse shown in Fig. \ref{rescaled}. We believe that the present paper achieves these tasks. 

In order to understand the broadband dielectric spectroscopy of glycerol (and in the future water-glycerol mixtures and other alcohols), we will treat glycerol in the relevant range of temperatures as a heterogeneous fluid on macroscopic timescales. That is, that while on very long timescales the liquid phase must be homogeneous, there exist localized mesoscale domains in the fluid that have macroscopic lifetimes. Indeed, inhomogeneities that appear to survive for $~10^4$ seconds contributing to the dielectric response in the Fourier domain at frequencies as low as $~10^{-4}$ Hz in some cases. We shall develop the theory on the basis that these inhomogeneities are a distribution of clusters having the structure of an incipient  strongly hydrogen-bonded phase and a surrounding bath of more mobile and less dense ``liquid-like" glycerol phase. Together therefore a snapshot of this phase
will present disordered "glass-like" characteristics. Our task is two-fold: First, to express the dielectric response in terms of this cluster distribution; and second, to find the distribution of clusters. We shall do this using mesoscale thermodynamic arguments. Once we have these distributions we describe the resulting BDS, and we shall see the appearance of both a dominant $``\alpha"$ peak at low frequencies and, depending on molecular parameters, an
`excess wing' or a secondary $``\beta"$ shoulder at higher frequencies. We should note that in spirit our approach
combines ideas based on the molecular dynamics observations by Geiger and Stanley~\cite{GS82} on the appearance of low-density patches in hydrogen-bonded water (though in the present context we are actually considering higher density patches in hydrogen-bonded glycerol), together with  ideas of Chamberlin \cite{93Cha}  and Kivelson et al. \cite{98KT,00TKV} on treating such heterogeneties as clusters. The similarity to the latter theories extends however only up to Eq. (\ref{response2}) below, after which our theory diverges from theirs.  The crucial differences from previous theories will be pointed out below as we go along. We trust that the reader will find the present approach superior in simplicity and in the quality of the predictions.

In Sect. \ref{clusters} we present the dielectric theory for a medium in which there exists clusters whose dipole moments
are responsible for the dielectric response. We estimate the relaxation time of such clusters, and most crucially,
their concentrations. In computing the concentrations we deviate most strongly from all previous theories. 
In Sect. \ref{glycerol} we specialize the theory to the case of glycerol; it is interesting to see how, by using a modest
input of experimental data we get naturally a host of results in very good correspondence with experiments. We explain the
apparent Vogel-Fulcher and stretched exponential fits, but also point out that the range of validity of these fits is finite, whereas the present theory is more widely applicable. Sec. \ref{beta} opens up the discussion for other materials,
and examines the changes expected in the spectra when molecular parameters change. We show how the relative prominence and $\alpha$ and $\beta$ peaks depend on such parameters. In Sect. \ref{pores} we discuss BDS in constrained geometries and
explain why glycerol in small pores exhibited the same spectra as bulk glycerol; the explanation is that the smallest
pores were just big enough to contain the largest clusters predicted by the theory. Reducing the pores a bit further
should result in a dramatic change in the spectra. Section \ref{discussion} presents a summary and a discussion. 

 \section{Dielectric Theory for Clusters}
 \label{clusters}
 \subsection{The Physical Model and the Fundamental Assumption}
 
We propose that the long-time relaxation in glycerol arises due to the existence of clusters of $s$ molecules which are hydrogen bonded to make the cluster a recognizable entity. We denote the number of clusters of $s$ molecules as $N_s$. In terms of these clusters we can write the dipole moment of the system, $\B M(t)$, as the sum ${\bf M}(t) = \sum_\alpha {\bf m}_\alpha (t) + {\bf M}_{liquid}(t)$ where ${\bf m}_\alpha (t)$ is the dipole moment of cluster $\alpha$ and the fast relaxing fluid contribution to the dipole moment ${\bf M}_{liquid}(t)$ will not be studied in detail  as its relaxation spectrum is expected to be significant at high frequencies. We expect that in the glassy state relaxation phenomena are relatively rare events, allowing us to assume that different relaxation events are statistically uncorrelated. This is
the fundamental assumption of the model, i.e.  that the relaxation process  in each cluster is statistically independent of the other clusters. In fact, by a ``cluster" we mean a bunch of molecules that are highly correlated; if there are two adjacent clusters that are highly correlated, they should be considered as one cluster. With this in mind we can then write
\begin{equation}
\label{response1}
\phi (t) = \frac{\sum_\alpha \langle {\bf m}_\alpha(t)\cdot{\bf m}_\alpha(0)\rangle}{\sum_\alpha \langle { \bf m}_\alpha(0)\cdot{\bf m}_\alpha(0)\rangle}.
\end{equation} 
We can rewrite Eq.~\ref{response1} in terms of an intensive number density  $n_s = N_s/M$ , where $M$ is the total number of molecules in the system: 
\begin{equation}
\label{response}
\phi (t) = \frac{\sum_s n_s \langle {\bf m}_s(t)\cdot{\bf m}_s(0)\rangle}{\sum_s n_s \langle { \bf m}_s(0)\cdot{\bf m}_s(0)\rangle}.
\end{equation} 
To each cluster we associate its longest typical relaxation time $\tau_s$ which we identify with the cluster lifetime. Thus $\phi (t)$ assumes the form
\begin{equation}
\label{response2}
\phi (t) = \frac{\sum_s n_s \langle {\bf m}_s\cdot{\bf m}_s\rangle \exp{(-t/\tau_s)}}{\sum_s n_s \langle { \bf m}_s\cdot{\bf m}_s\rangle}.
\end{equation} 
To proceed further we need to estimate $n_s$, $\langle {\bf m}_s\cdot{\bf m}_s\rangle$ and $\tau_s$. We start with the static dipole correlations. 
\subsection{The static dipole correlations}

The dipole moment of a cluster of size s can be expressed in terms of individual glycerol molecule dipoles ${\bf d}_i$ as ${\bf m}_s = \sum_{i=1}^s {\bf d}_i$. For a set of non-interacting dipoles $ \langle {\bf m}_s\cdot{\bf m}_s\rangle = s d^2$, where $d=|{\bf d}_i|$. We do not expect the dipoles in the clusters to be totally random, but rather to exhibit strong {\bf short-range} correlations due to dipole-dipole interactions. This short range order was taken into account explicitly by
Kirkwood \cite{39Kir} who introduced the so-called Kirkwood factor $g$: $ \langle {\bf m}_s\cdot{\bf m}_s\rangle = gs d^2$, where $g>1$. Obviously this constant is difficult to calculate for a given material, but luckily it appears in both the numerator and denominator of $\phi (t)$ and consequently Eq.~(\ref{response2}) reduces to the expression
\begin{equation}
\label{response3}
\phi (t) = \frac{\sum_s n_s s \exp{(-t/\tau_s)}}{\sum_s n_s s}.
\end{equation} 
If we insert Eq.~\ref{response3} into Eq.~\ref{diel} and split the dielectric constant into its real and imaginary parts $\epsilon (\omega) = \Re  \epsilon (\omega) + i \Im \epsilon (\omega)$ we find
\begin{eqnarray}
\label{result}
\frac{\Re \epsilon (\omega ) - \epsilon_{\infty}}{(\epsilon_0 - \epsilon_{\infty}) }& =&  \frac{\sum_s n_s s/[1 + (\omega \tau_s)^2] }{\sum_s n_s s} \\
\frac{\Im \epsilon (\omega ) }{(\epsilon_0 - \epsilon_{\infty}) }& = &  \frac{\sum_s n_s s (\omega \tau_s)/[1 + (\omega \tau_s)^2] }{\sum_s n_s s} \nonumber\\  &+& \frac{4\pi  \sigma_{dc}/(\epsilon_0 - \epsilon_{\infty} )}{\omega} \  . \nonumber
\end{eqnarray}
Thus we see that we have expressed both the broadband dielectric constant and loss in terms of the cluster size distribution and the cluster lifetimes. 
\subsection{Relaxation times $\tau_s$}

The relaxation time of the clusters, mainly due to their rotational relaxation, will be determined by the free energy barrier that involves breaking the hydrogen bonds with the surrounding liquid.
We argue that these are given by Arrhenius forms where the energy barrier scales with the
surface area of the cluster as the clusters attempts to break the cage of mobile 'liquid-like' molecules in which it is confined.
\begin{equation}
\label{time}
\tau_s = \tau_0 \exp{(\bar{\mu} s^{2/3}/k_BT)} \ .
\end{equation}
Here the attempt time $\tau_0$ is of the order of $10^{-12}$ seconds; while the energy for breaking a typical bond  $\bar{\mu}\approx \sigma$ can be expected
to scale with the surface energy per molecule. We will estimate the numerical values of these
parameters below in the context of the theory for glycerol.

While Eq. (\ref{time}) appears rather innocent and perfectly reasonable, its consequences are manifold, lying at the
very basis of our approach. These consequences are explained in the next subsection.

 \subsection{Mesoscale Thermodynamic Theory for the Cluster Distribution $n_s(T,p)$.}
 
 The most interesting part of the theory is the estimate of the cluster size distribution $n_s(T,p)$.
In this section we provide a thermodynamic theory to estimate this crucial quantity. Obviously, any such approach has to face the fact that the glassy state is thermodynamically {\bf metastable}, with the ground state always being the crystalline solid, i.e. a single ``cluster" with $s=\infty$. The actual nature of this metastable state depends on the preparation protocol.
As is well known, different rate of cooling for example can end up in glassy states where the properties
of the metastable state are different. Moreover, left alone, such a state will  ``age" \cite{00SBLL}, changing ever so intermittently and ever so slowly towards the ground state. What we are really talking about is dynamical arrest. 

To see the role of dynamics, note that if the relaxation time as expressed in Eq. (\ref{time}) is correct we can estimate the size of the largest clusters as $s_{max}$ just from looking at the longest relaxation times observed experimentally. 
We thus write
\begin{equation}
\label{estsmax}
s_{max}  \sim [(k_BT/\bar{\mu})\ln (t_{max}/\tau_0)]^{3/2}
\end{equation}
 where $t_{max}$ is the longest time scale observed in the broadband spectrum. For example, for glycerol $t_{max}\sim 10^4$ seconds and $\bar{\mu}\sim 2 k_B T$;  then $s_{max}\sim 100$. Of course, this number can
 depend on the cooling protocol, and when the system ages we expect $s_{max}$ to be a slowly (and probably intermittently) increasing function  of time. Nevertheless we will see below that in {\bf order of magnitude} this estimate is fully justified by the present theory. But more importantly, we can now turn the argument around, and say that if the cluster distribution is stationary, then the relaxation time of the cluster should be of the same order as its creation time. So let us estimate, if it takes $10^4$ seconds to create a cluster with $s_{max}=100$, how long will it take to create a cluster of size $2s_{max}$. Using Eq. (\ref{estsmax}) with $\tau_0\approx 10^{-12}$ yields the estimated time to create
 a cluster of 200 molecules to be longer than $10^{13}$ seconds! This is the crux of the matter. The dynamical arrest
 can be interpreted here as having a distribution of clusters in which $s_{max}$ is sharply determined by the aging
 time of our glassy state. In the language of phase transitions we can say that the phase transition had occurred,
 but the coarsening of the cluster distribution is inhibited by dynamical arrest. In any reasonable theory of
 coarsening, like, say, the Lifshitz-Slyozov theory \cite{61LS}, the time scale is determined by the diffusion coefficient, which in our super-cooled liquid is many orders of magnitude smaller than in regular liquids undergoing a phase transition. Thus at any time we have a maximal cluster, and the smaller clusters which form much quicker have had ample time to equilibrate according to the regular laws of thermodynamics. 
 
 Thus, In developing a thermodynamic theory,
we need to invoke a constraint on the size of the maximal cluster. In other words, we treat the heterogeneities in glycerol in terms of a mesoscale distribution of clusters with a range of size $s_{min}< s < s_{max}$ in a more mobile bath of  `liquid-like'  molecules.  This approach of applying to the glass-forming liquid statistical
 mechanics subject to constraints follows up our previous analysis of other glass forming liquids, cf. \cite{07EBIMPS,07HMPS,07ILLP}. Note that this approach is radically different from the proposition of \cite{98KT,00TKV} where a mysterious ``strain energy" term was invoked to explain why larger clusters are not present in the theory.
 
 The intensive number of such clusters of size $s$ are $n_s(p,T) = N_s(p,T)/M$, and the probability that an individual glycerol molecule belongs to a cluster of size $s$ is $c_s = n_s s$. 
 We denote below the number of molecules belonging to the clusters and to the liquid phase by
 $M_c$ and $M_\ell$ respectively, with $M_c+M_\ell=M$. To describe the mesoscale thermodynamics we assume that  in a cluster at temperature $T$ and pressure $p$ we have an intensive contribution to the chemical potential 
\begin{equation}
\label{mui} 
 \mu_c = u_c+ p v_c -T s_c, 
 \end{equation}
 where the subscript $c$ stands for ``cluster" and $u_c$, $v_c$, $s_c$ are respectively the internal energy/molecule, the volume/molecule, and the entropy/molecule in the clusters. Similarly in the `liquid-like' phase we can write for the intensive contribution to the chemical potential  
 \begin{equation}
\label{mu} 
 \mu_\ell = u_\ell + p v_\ell -T s_\ell, 
 \end{equation}
where the subscript $\ell$ stands for liquid and $u_\ell$, $v_\ell$, $s_\ell$ are respectively the internal energy/molecule, the volume/molecule, and the entropy/molecule in the mobile 'liquid-like' phase. We expect that $u_c< u_\ell$ because of the the hydrogen bonds in the clusters are less distorted; $s_c < s_\ell$ because of the greater number of rotational degrees of freedom in the mobile phase; and  $v_c < v_\ell$ because of the higher density of more solid-like glycerol compared to the liquid. All these intensive thermodynamic variables are in principle functions of temperature T and pressure p. Apart from these extensive contributions to the Gibbs free energy $M_c \mu_c + M_\ell \mu_\ell$ where $M_c  = \sum_s N_s s = M \sum_s n_s s = M \sum_s c_s$, we need to consider two other crucial contributions to the total free energy. 
 
 First, there is a surface energy contribution $U_{surface}$ due to the interface between each cluster and its surrounding bath. This scales as the surface area of each cluster and therefore we can write
 \begin{equation}
 U_{surface} = \sigma \sum_s N_s s^{2/3},
\end{equation}
where $\sigma$ is the surface energy/molecule. Obviously, the value of $\sigma$ will turn out to be crucial in determining the distribution of cluster sizes. To estimate its size note that $\sigma$ could be as large as a hydrogen bond energy, but in fact we expect it to be smaller because the mobile phase
is also partially hydrogen bonded.  Thus we expect $\sigma \approx u_\ell - u_c$ which is some fraction of a hydrogen bond energy.

Second, there will be an entropic contribution due to all possible spatial configurations of clusters in the glassy liquid.  This is an important contribution to the free energy, {\bf favoring small clusters and being responsible for the $\beta$ peak when it exists}. In taking the entropy explicitly into account we deviate once more from all previous theories; we cannot see why the entropy was not considered before. We can estimate this entropy by sequentially placing all clusters in the available volume starting with the largest \cite{07HMPS}.
Thus the volume available to the largest clusters is  $V = M_c v_c  + M_\ell v_\ell$. The volume available to the next largest clusters is $V - s_{max}N_{s_{max}}v_c$ etc. Proceeding
in this manner the entropy of mixing is given by the expression
\begin{equation}
\label{entropy}
S_{mix} = - k_B \sum_s \mathcal{N}_s [ x_s \ln x_s + (1-x_s) \ln (1- x_s)]
\end{equation}
where $k_B$ is Boltzmann's constant and $\mathcal{N}_s = (V/v_c - \sum_{s'>s}N_{s'}s')/s$ is the number of boxes available to clusters of size $s$ while $x_s = N_s/\mathcal{N}_s$ are the fraction that are occupied. Thus
\begin{equation}
\label{x}
x_s = \frac{c_s }{\sum_{s'=s_{min}}^{s} c_{s'} + (v_\ell/v_c)c_\ell}.
\end{equation}
where $c_\ell$ is the fraction of molecules in the 'liquid-like' phase, and of course
the sum of the concentrations of molecules in the "condensed' and 'liquid-like' phases obey 
\begin{equation}
c_c+ c_\ell=1.0 \ . \label{sum}
\end{equation}

At this point we introduce $v$ as the average volume per glycerol molecule in our system, where of course 
\begin{equation}
v= c_c v_c + c_\ell v_\ell \ . \label{sum1}
\end{equation}
We can now solve Eqs. (\ref{sum}) and (\ref{sum1}) for $c_c$ and $c_\ell$ in terms of these
3 volumes per molecule and write the crucial sum-rule:
\begin{equation}
\label{constraint}
\sum_{s = s_{min}}^{s_{max}} c_s \equiv c_c = \frac{v-v_\ell}{v_c - v_\ell} \ .
\end{equation}
This is the crucial constraint  on the thermodynamic theory ; we will  impose this constraint on our solution by choosing the largest cluster $s = s_{max}$ such that Eq.~(\ref{constraint}) is obeyed.

The Gibbs free energy for our system is given by the expression
\begin{eqnarray}
\label{G}
&&G = U + PV -TS = M_c \mu_c + M_\ell \mu_\ell + \sigma \sum_s N_s s^{2/3}\nonumber\\
&&+k_BT \sum_s \mathcal{N}_s [ x_s \ln x_s + (1-x_s) \ln (1- x_s)].
\end{eqnarray}
and the chemical potential of a cluster of size $s$ is given by
\begin{eqnarray}
\label{mus}
&&\mu_s = s \mu = \partial G/\partial N_s = \mu_c s + \sigma s^{2/3}\nonumber\\ &&+ k_BT\{ \ln \psi_s + \sum_{s' > s} (s/s') \ln (1 - \psi_{s'}) \}
\end{eqnarray}
where
\begin{eqnarray}
\label{psi}
\psi_s &=& \frac{c_s}{\sum_{s'= s_{min}}^{s} c_{s'} + \Delta } \ , \\
\Delta &=& \frac{v_\ell}{v_c}\frac{v_c - v}{v_c - v_\ell} \nonumber\ .
\end{eqnarray}
Note that $0< \Delta  <1$. Rewriting Eq.~\ref{mus} for the chemical potential per  molecule in cluster
\begin{equation}
\label{mus2}
\mu = \mu_c  + \sigma s^{-1/3} + k_B T\{ (1/s)\ln \psi_s + \sum_{s' > s} (1/s') \ln (1 - \psi_{s'}) \}
\end{equation}
The chemical potential in the liquid phase must be the  same as that calculated for the clusters and given by Eq.~\ref{mus2},
and therefore
\begin{equation}
\label{mus3}
\mu = \mu_\ell + k_BT (v_\ell/v_c)\{ (1/s)\ln \psi_s + \sum_s (1/s) \ln (1 - \psi_{s}) \}.
\end{equation}
We can use Eq.~(\ref{mus3}) to define a new chemical potential $\mu' = \mu - (\mu - \mu_\ell)/(v_\ell/v_c) - \mu_c$ which together with Eq.~(\ref{mus2}) then obeys
\begin{eqnarray}
\mu' &=&  \sigma s^{-1/3} + kT\{ (1/s)\ln [ \psi_s/(1.0-\psi_s)] \nonumber\\
&-& \sum_{s' < s} (1/s') \ln (1 - \psi_{s'}) \} \ . \label{mus4}
\end{eqnarray}
In this form it is easy to solve these equations numerically in a sequential fashion starting with $s = s_{min}$. Using Eq.~(\ref{psi}) we see that the concentration
of the smallest clusters $c_{s_{min}}$ obeys

\begin{equation}
\label{mus5}
\mu' =  \sigma s_{min}^{-1/3} + k_B T (1/s_{min})\ln [ c_{s_{min}}/\Delta ] .
\end{equation}
To estimate $s_{min}$ we note that in 3-dimensions a `cluster' must have at least 2-3 molecules in each direction to be identifiable as a cluster. Thus the minimal value of $s$ is about 20.
We do not know $\mu'$ {\it a priori} depending as it does on the details of the energies, structure and entropies in the clusters and 'liquid-like' phases, so instead we choose the reasonable values for $\mu' \sim k_B T$. We are now in a position to estimate $c_{s_{min}}$ from Eq.~\ref{mus5}.
Then all the values of $c_s$ for $s>s_{min}$ follow from Eqs.~(\ref{mus4}) and~(\ref{psi}) in order until Eq.(~\ref{constraint}) is fullfilled. We will see below that for glycerol the largest $s$ predicted\
by the theory is about 100.

\section{Predictions for Broadband Dielectric Spectroscopy: the case of Glycerol}
\label{glycerol}

Examining our expressions for the distribution of clusters given by Eqs.~(\ref{mus4})-(\ref{mus5}), we note that in order to apply the theory to a particular substance, we miss the value of the surface energy $\sigma$, say for glycerol. Since this quantity appears inside exponential forms we need a rather accurate value for comparison with experimental data. One approach would be to fit our dielectric spectra at one temperature and then use the fitted parameters to predict its form at different
temperatures. Here we have taken a slightly different approach, using available glycerol data. The are two types of data; the first concerns the average volumes per glycerol molecule in the liquid, solid and glassy phases. 
For $v_\ell$  we employed data of liquid glycerol at room temperature giving $v_\ell =121 \AA^3$. For $v_c$ and
$v$ we took data from \cite{89RS} 
which give for the
the cluster phase $v_c=110 \AA^3$ and for the glassy phase $v=113 \AA^3$ at atmospheric pressure. These
numbers were used to estimate the constraint given by Eq.~(\ref{constraint}). The second type of data we have at our disposal concerns the size of dynamic heterogeneities as a function of temperature. We have taken data from \cite{07LTL} for the average size of such heterogeneities $s_{av} = \sum_s s c_s/\sum_s c_s$. The data used were $s_{av} (T=210K)\approx 50$,
$s_{av} (T=205K)\approx 58$, $s_{av} (T=200K)\approx 67$, $s_{av} (T=195K)\approx 73$. These
two pieces of information can be accommodated in our theory  using $\sigma /k_B = 320K$ or $\sigma /k_B T \approx 1.6$. We then predict the maximal cluster sizes at these temperatures to be $s_{max} (T=210K)\approx 63$, $s_{max} (T=205K)\approx 72$, $s_{max} (T=200K)\approx 81$, $s_{max} (T=195K)\approx 87$. Using this data we then studied the theoretical predictions for glycerol.

\begin{figure}
\centering
\epsfig{width=.7\textwidth,file=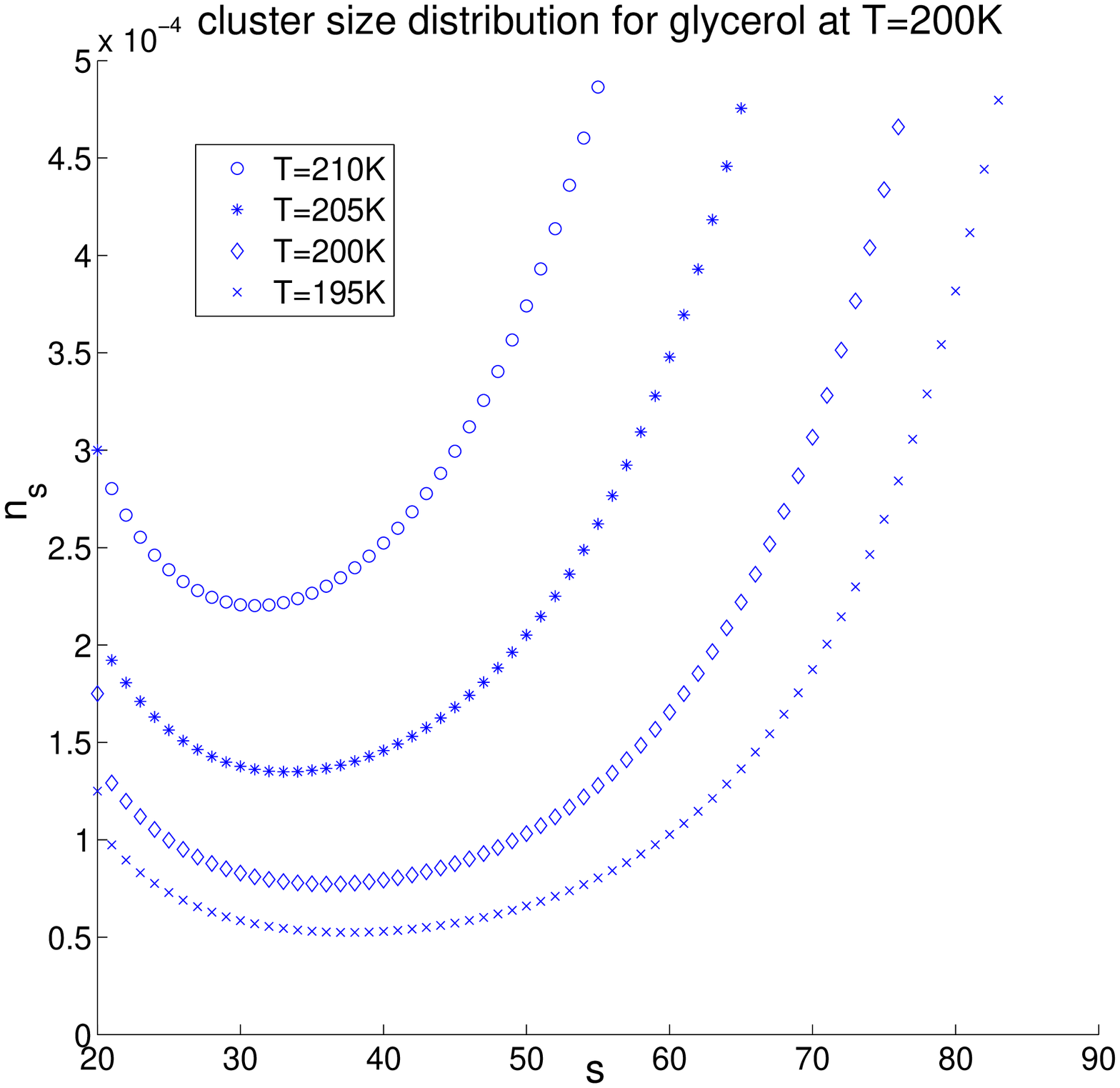}
\epsfig{width=.7\textwidth,file=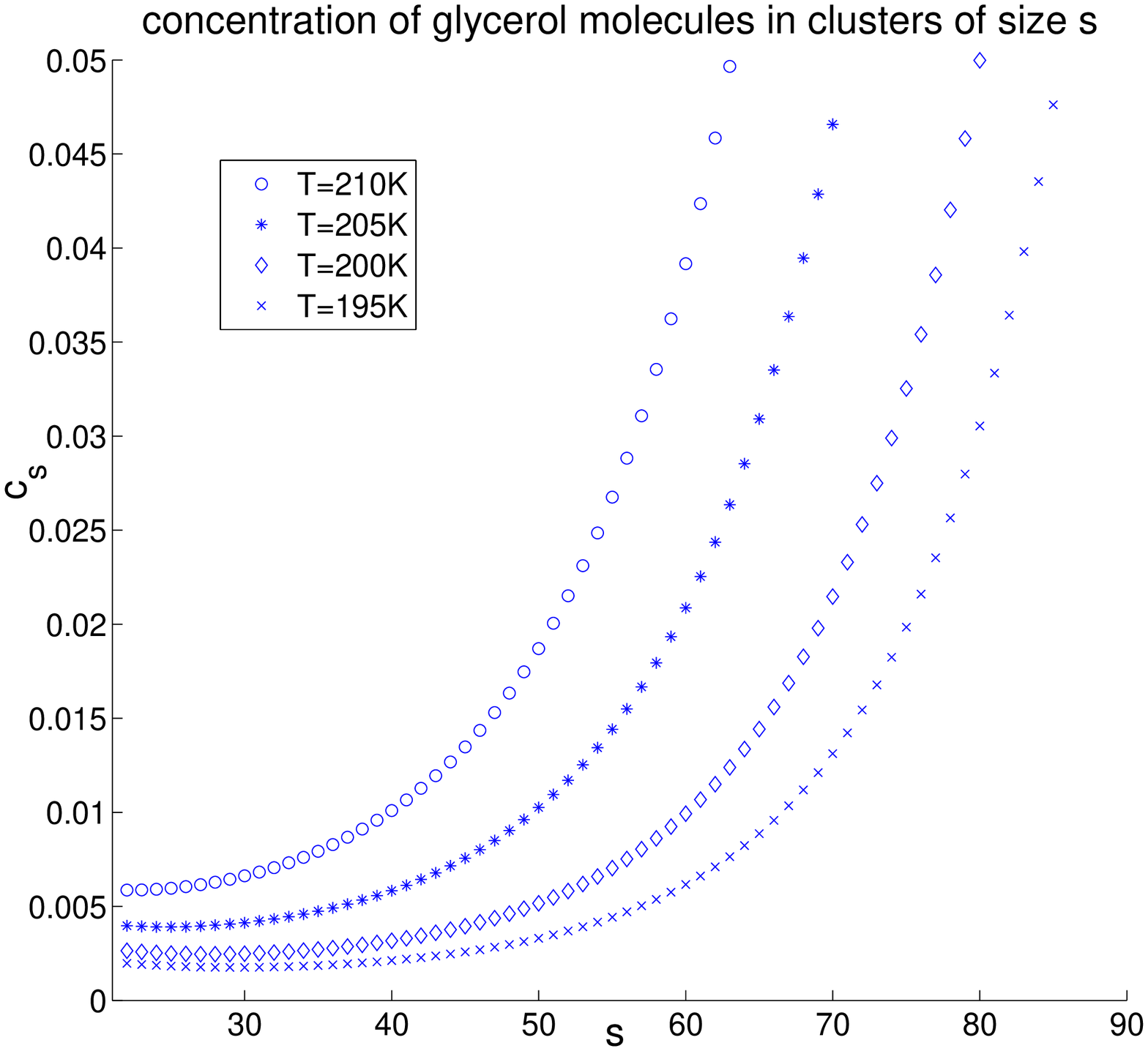}
\caption{Upper panel: Cluster size distribution $n_s$ versus $s$ for temperatures $T=210K, 205K, 200K,195K$. Lower panel: Concentration of molecules $c_s$ as a function of cluster size $s$ for temperatures$T=210K, 205K, 200K,195K$.}
\label{NC}
\end{figure}

As noted above, the distribution of clusters between $s_{min}$ and $s_{max}$ is strongly dependent on surface energy. Using $\sigma/k_B = 320K$, we plot the cluster size distribution $n_s$ and the density of molecules as a function of cluster size $s$  for dry glycerol at $T= 210K, 205K, 200K, 195K$. Small clusters are favoured at higher temperatures,  when the entropy of mixing dominates the distribution (see Fig.~\ref{NC}). But as the temperature is reduced larger clusters are favoured as energy dominates the distribution. In general the distribution is bimodal favouring small and large clusters at the expense of intermediate sizes (see Fig.~\ref{NC}). We note that the sharp cut of the distribution at $s=s_{max}$ is
a bit artificial; in reality one can expect a sharply decaying tail at $s$ values slightly larger than $s_{max}$. We do not
expect such minor details to influence the main results presented below.

\begin{figure}
\centering
\epsfig{width=.5\textwidth,file=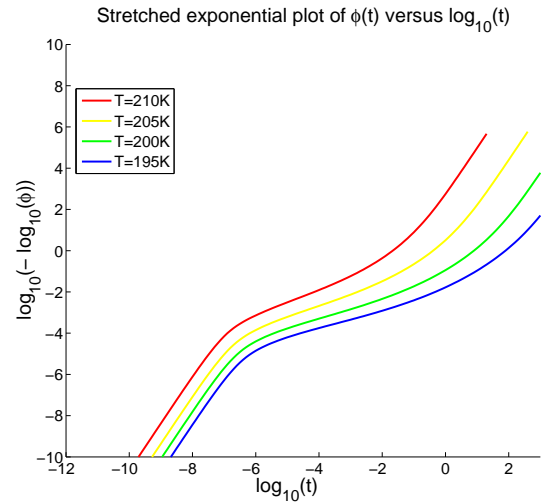}
\caption{ Stretched exponential plot of $\phi (t)$ showing and approximate straight line over about six orders of magnitude in time for temperatures $T=210K$ (leftmost), 205K, 200K, 195K (rightmost).}
\label{Stretch}
\end{figure}

We are now in a position to calculate both $\phi (t)$ from Eq.~(\ref{response3}) and the real and imaginary parts of the dielectric function from Eq.~(\ref{result}). This quantity had been fitted phenomenologically to a stretched exponential form, $\phi (t) \sim \exp{- (t/\tau)^{\beta_K}}$, 
a form referred to in the literature as the ``Kohlrausch-Williams-Watts (KWW) relaxation function". 
(cf. for example \cite{06HPF}).  To see  whether this form is justified by the present theory we plot our computed function in the appropriate coordinates, cf.  Fig.~\ref{Stretch}. Indeed, a stretched exponential with $0.4<\beta_K <0.66$ gives an acceptable fit over a broad range of timescales $t_{min}\sim 10^{-6} \ll t \ll t_{max}\sim 10^0$ secs, with deviations  at both shorter and longer times. We note that $\beta_K$ is temperature dependent and stress that the stretched exponential form has a limited value in the sense
that it is just an acceptable fit in a limited range for a very different function. Nevertheless 
the numerical value of $\beta_K$ and its variability with the experimental conditions are both confirmed by experiments in glycerol, see for example \cite{05PHRBFKB,06HPF}. 

\begin{figure}
\centering
\epsfig{width=.6\textwidth,file=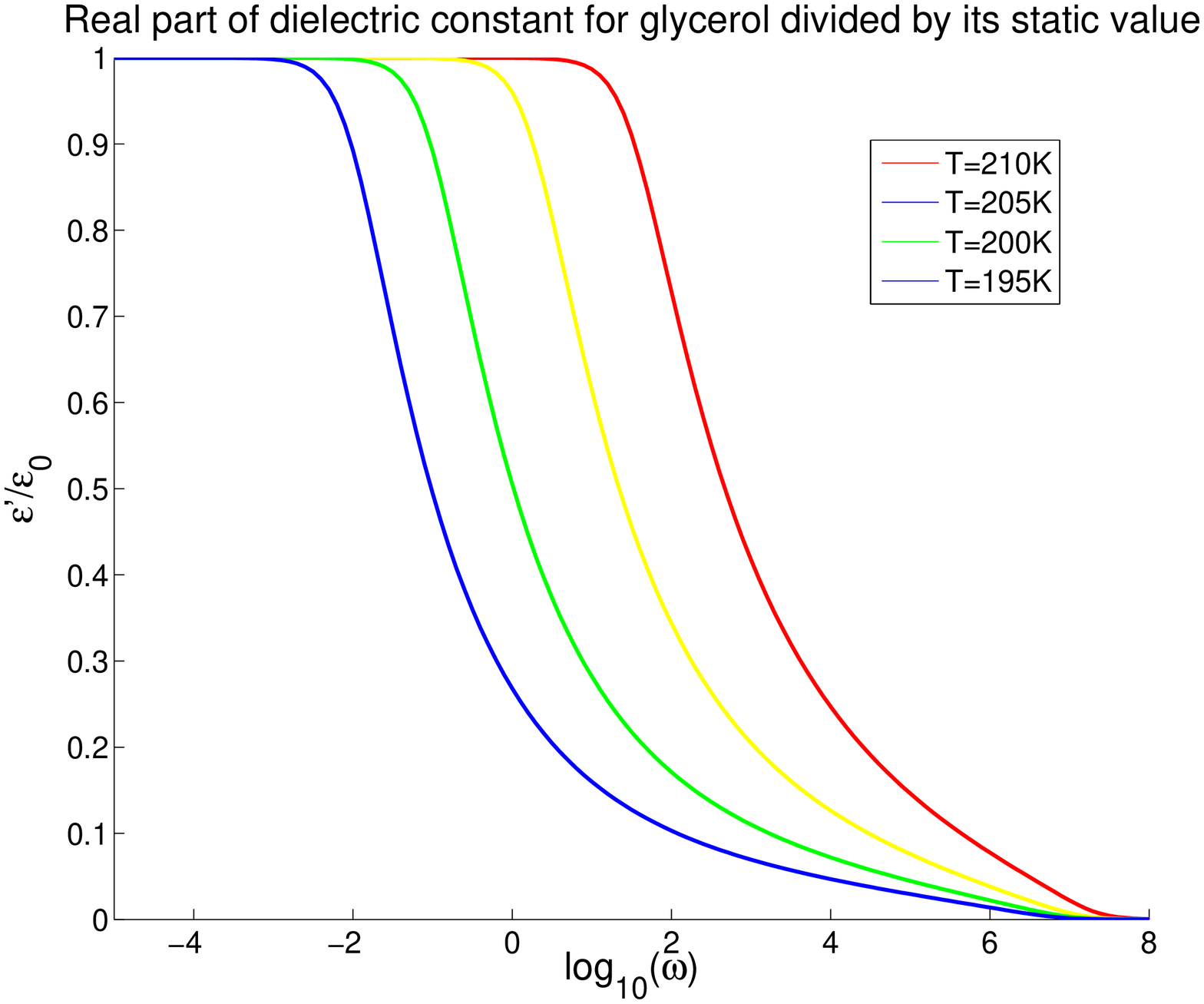}
\epsfig{width=.6\textwidth,file=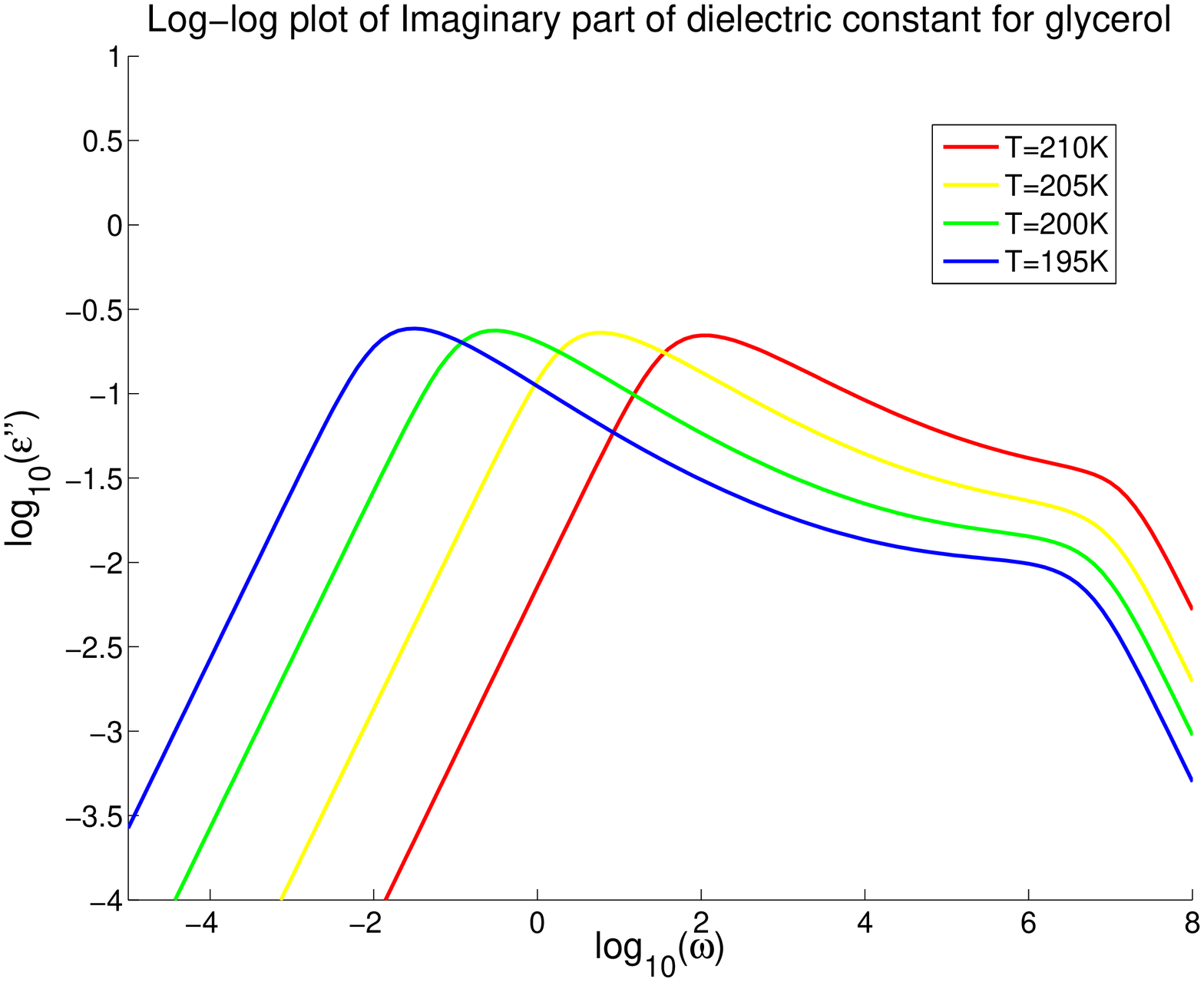}
\caption{Upper panel: Real part of the dielectric spectrum $\epsilon'(\omega )/\epsilon_0$ versus  $\log_{10} (\omega )$ for temperatures $T=210K,~205K,~200K,~195K$. Lower panel: A log-log plot of the dielectric loss $\log_{10}(\epsilon''(\omega ))$ versus  $\log_{10} (\omega )$ for temperatures $T=210K$ (rightmost), 205K, 200K and 195K (leftmost)}
\label{EPSILON}
\end{figure}

Of greater interest is the  dielectric spectrum and loss function (see Fig.~\ref{EPSILON}) which correspond to the distributions shown in Fig.~\ref{NC} with $T= 210K,205K,200K,195K$ respectively). 
The real part of the dielectric constant $\Re \epsilon (\omega )$ start to decline from $\epsilon_0$ to
its asymptotic value of $\epsilon_{\infty}$ at around $\omega \approx \omega_{max}$ but takes several
frequency decades to achieve asymptotia. We refer the reader to \cite{05PHRBFKB,06HPF} and references therein to note that both the qualitative form of $\Re \epsilon (\omega )$and its quantitative
details are in close correspondence with experiments in glycerol (as well as in glycerol-rich water mixtures).

The  shape  of $\Im \epsilon (\omega )$ is controlled by the shape of the cluster size distribution. There is always a clear $\alpha$ peak at low frequencies $\omega_{max}$ defined by $d\Im \epsilon (\omega_{max} )/d\omega_{max} = 0$ associated with the largest clusters. Very roughly $\omega_{max} \sim 1/t_{max}$. 
But we can also see at all temperatures a clear  `excess wing'  at higher frequencies. As  the temperature is lowered the `excess wing'  becomes slightly  less pronounced. 

\begin{figure}
\centering
\epsfig{width=.5\textwidth,file=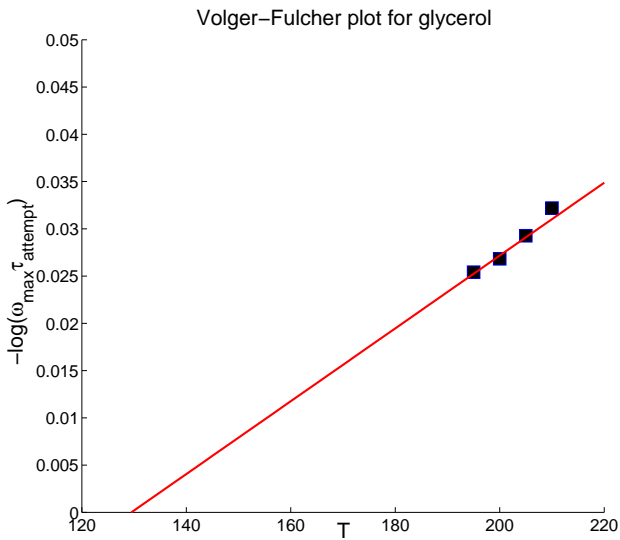}
\caption{$\omega_{max}$ for the dielectric loss  for temperatures $T=210K, 205K, 200K, 195K$ plotted in a Vogel-Fulcher form. The plot underlines the irrelevance of $T_v$ as a physical parameter, since it is so far removed from the region of linear fit. The theory presented here actually does not predict a true divergence of the relaxation time, it only becomes super-exponentially large. }
\label{Vogel}
\end{figure}

The next interesting question is whether we can justify theoretically the Vogel-Fulcher plots. To answer
this question we plot the computed values of $\omega_{max}$ in a Vogel-Fulcher form as in Eq.~(\ref{VF}), cf. Fig~\ref{Vogel}.
In this plot we display, in purpose, all the temperature range including $T_v$ to stress the absurdity of 
such a fit. Nevertheless a straight line can be fitted through the computed points.  Using the same (unphysical) attempt time of $\ln\tau_v=-35.9$ as in the experimental fits on glycerol the straight line
best fit gives $T_v = 129K$ and a fragility $D=20.0$. This should be compared with $T_v=125K$ and  $D=22.7$ which are the numbers reported experimentally. Our conclusion is that the theory explains both the stretched exponential fits to the relaxation function and the Vogel-Fulcher fit, but both are fundamentally meaningless, and should be replaced by a theory of the type proposed here.

\section{The role of Surface Energy: Cluster Distributions and Dielectric Spectra}
\label{beta}

\begin{figure}
\centering
\epsfig{width=.5\textwidth,file=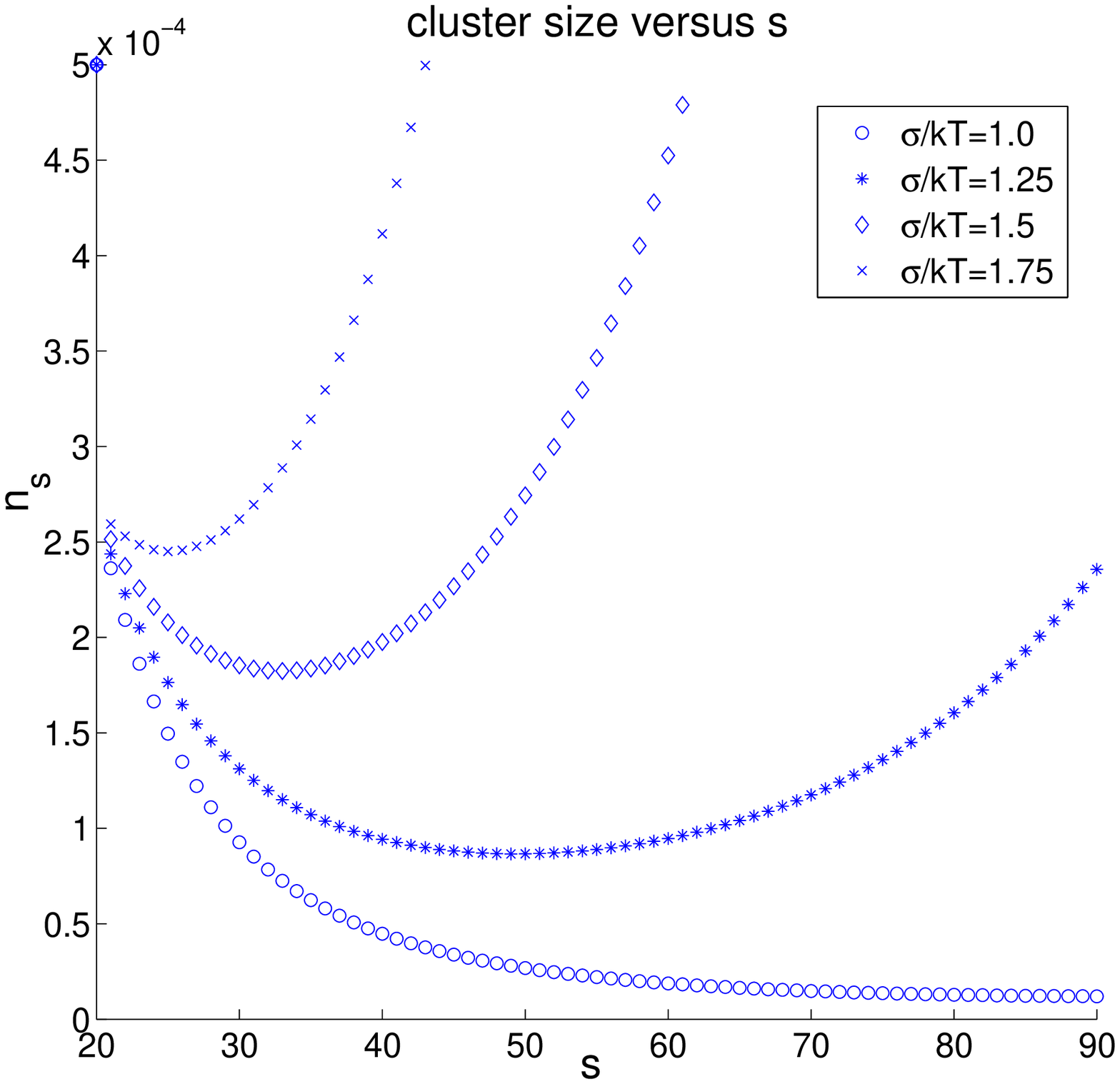}
\epsfig{width=.5\textwidth,file=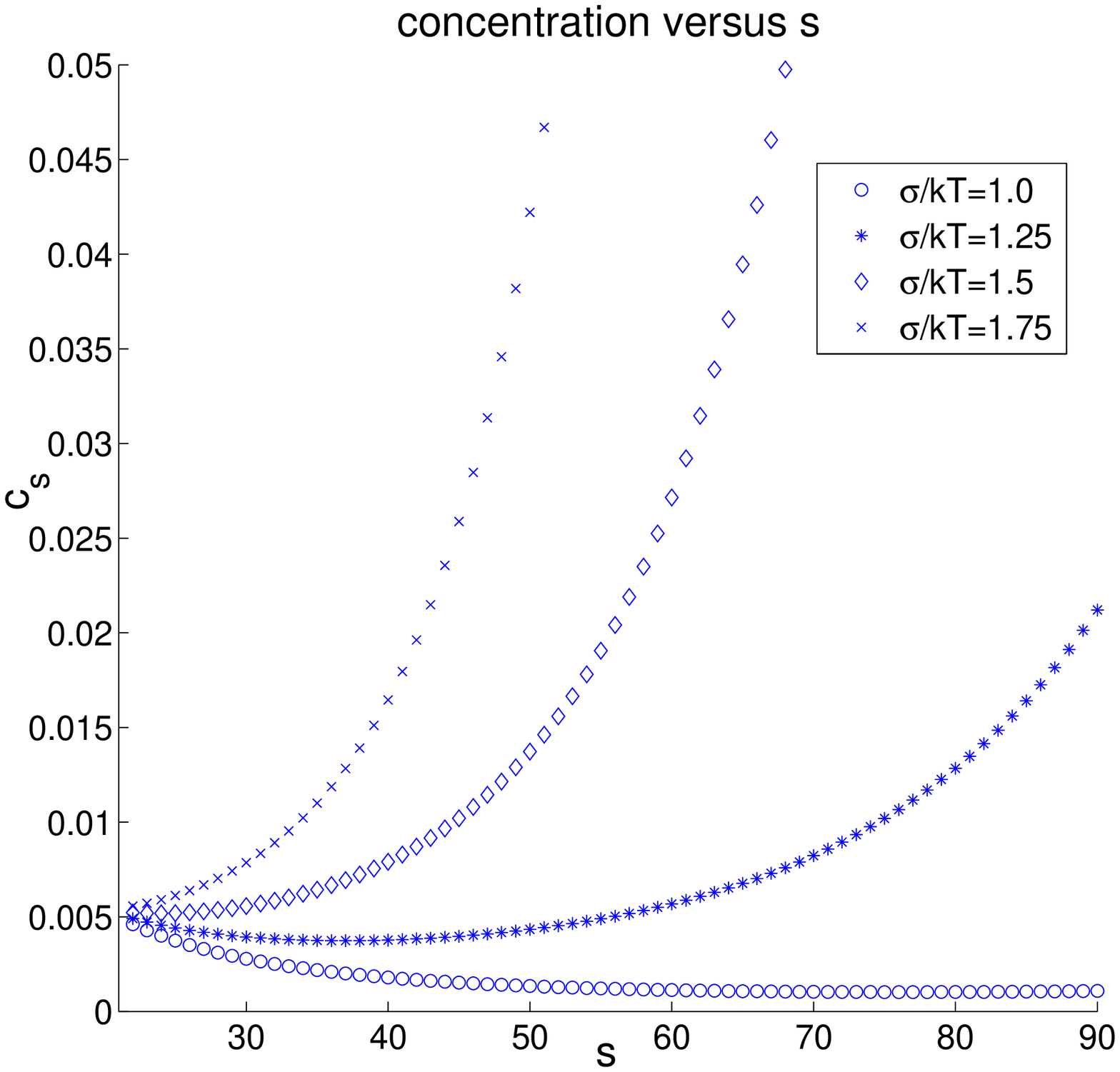}
\caption{Upper panel: Cluster size distribution $n_s$ versus $s$ for $\sigma/k_B T =1.0, \sigma/k_B T =1.25, \sigma/k_B T =1.5, \sigma/k_B T =1.75$. Lower panel: Concentration of molecules $c_s$ as a function of cluster size $s$ for temperatures  $\sigma/k_B T =1.0, \sigma/k_B T =1.25, \sigma/k_B T =1.5, \sigma/k_B T =1.75$.}
\label{NC2}
\end{figure}

To understand the experimental results for glycerol we chose the crucial parameter $\sigma$ from structural data. In this section we ask a different question --- what is the qualitative influence of the surface energy $\sigma$ on cluster size distributions and dielectric responses. In particular we are interested in the possibility of generating a distinct
$\beta$ peak by changing only one molecular parmater, which is $\sigma$. To study this question we fix $s_{max}=100$  and $T=200K$ and study the effect of changing the surface energy.
We choose four values $\sigma/k_B T =1.0,~1.25,~1.5$ and 1.75. As we increase $\sigma$ the average cluster size increases:  $s_{av} (\sigma/k_B T =1.0)\approx 48$,  $s_{av} (\sigma/k_B T =1.25)\approx 76$, $s_{av} (\sigma/k_B T =1.5)\approx 87$, $s_{av} (\sigma/k_B T =1.75)\approx 91$. The complete distribution can be seen in Fig.~\ref{NC2}.
As $\sigma$ increases large clusters are favoured over small clusters; $\sigma$ has a crucial  qualitative influence on the bi-modality  of the distribution. 

\begin{figure}
\centering
\epsfig{width=.5\textwidth,file=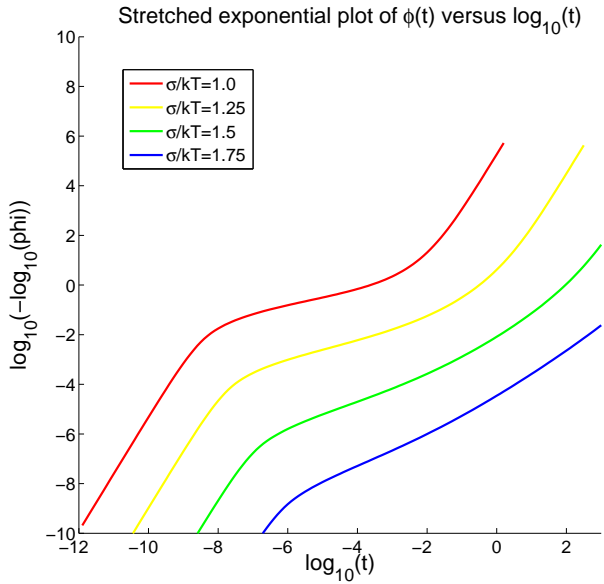}
\caption{ Stretched exponential plot of $\phi (t)$ to show scaling regime  in time for $\sigma/k_B T =1.0, \sigma/k_B T =1.25, \sigma/k_B T =1.5, \sigma/k_B T =1.75$.}
\label{Stretch2}
\end{figure}

The change in surface energy must also influence the dynamics. We asserted that the lifetimes $\tau_s$ of clusters of size s are given by Arrhenius forms where the energy barrier scales with the
surface area of the cluster, as the clusters attempts to break the cage of mobile 'liquid-like' molecules in which it is confined. The energy for breaking a typical bond  can be expected
to scale with the surface energy $\sigma$. In Fig.~\ref{Stretch2} we have plotted the behaviour of $\phi (t)$ in a stretched exponential form to show scaling  in time for $\sigma/k_B T =1.0,~1.25,~1.5$ and 1.75.
Note that as $\sigma$ increases the stretched exponential regime increases due to a more pronounced $\alpha$ peak while at low surface energies the high frequency $\beta$ peak destroys this kind of
scaling.

\begin{figure}
\centering
\epsfig{width=.5\textwidth,file=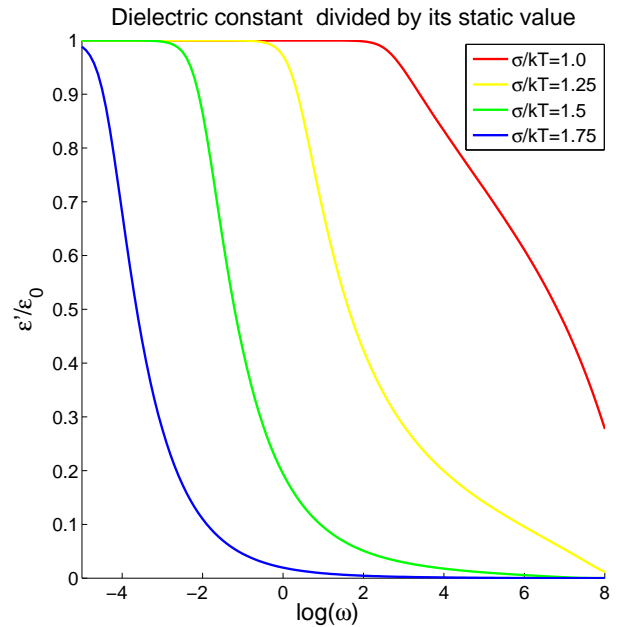}
\epsfig{width=.5\textwidth,file=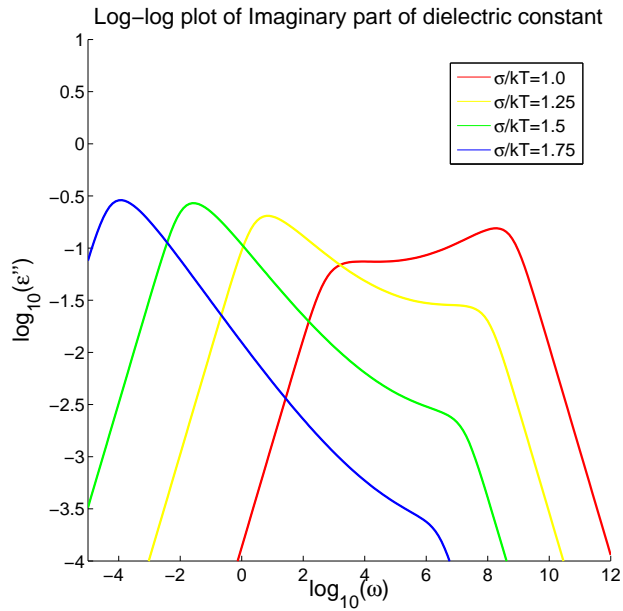}
\caption{Upper panel: Real part of the dielectric spectrum $\epsilon'(\omega )/\epsilon_0$ versus  $\log_{10} (\omega )$ for $\sigma/k_B T =1.0$ (rightmost),  $\sigma/k_B T =1.25$,  $\sigma/k_B T =1.5$, $\sigma/k_B T =1.75$ (leftmost).  Lower panel: A log-log plot of the dielectric loss $\log_{10}(\epsilon''(\omega ))$ versus  $\log_{10} (\omega )$ for the same values of $\sigma/k_B T$ as in the upper panel.}
\label{EPSILON2}
\end{figure}

Of greater interest are the  dielectric spectra and loss functions (see Fig.~\ref{EPSILON2}) which correspond to the distributions shown in Fig.~\ref{NC2} with  $\sigma/k_B T =1.0,~1.25,~1.5$ and 1.75 respectively). The shape of $\Re \epsilon (\omega )$ corresponds to a decline from $\epsilon_0$ to $\epsilon_{\infty}$ starting at $\omega_{max}$, but
one interesting point is that the $\beta$ peak can lead to a shoulder here also.
The  shape  of $\Im \epsilon (\omega )$ is again controlled by the cluster size distribution. There is always a clear $\alpha$ peak at low frequencies.
But for lower values of the surface energy small clusters are encouraged resulting in a prominent $\beta$ peak  at high frequencies. As the surface energy increases  the $\beta$ peak becomes less pronounced turning first into a shoulder at intermediate frequencies (see Fig.~\ref{EPSILON2}), and finally into the anomalous scaling observed in many experimental data sets (see Fig.~\ref{EPSILON2}).
These interesting qualitative findings will be turned to quantitative comparisons with experiments in different materials in a later publication.

\section{BDS in Confined Geometries}
\label{pores}

Finally, we discuss the BDS of glycerol in confined geometries. We refer here in particular to the experimental studies reported in \cite{96ASGK}, in which it was shown that glycerol could be confined in pores whose diameter $d$ can be as small as $d=2.5$ nm in size without seeing any appreciable effect in the position of the $\alpha$
maximum or its amplitude. Since we expect that the maximum size cluster should be limited by the size of pores, it appears surprising that there is no confinement effect on the spectra. 
If our arguments are correct as presented in this paper, there must be a point at which the dielectric loss will be strongly effected by confinement. To estimate where this should happen we equate the volume of the pore to the volume of the largest cluster, i.e.
\begin{equation}
\pi d^3/6 \approx s_{max}v_c \ .
\end{equation}
\begin{figure}
\centering
\epsfig{width=.5\textwidth,file=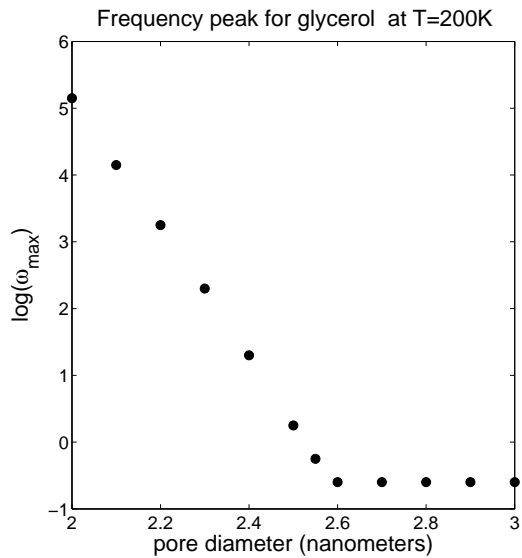}
\caption{Shift in expected position of $\omega_{max}$ as a function of confining pore diameter.
Note that no effect appears until a critical pore diameter after which the shift is dramatic.}
\label{PORE}
\end{figure}
Using our estimates of $v_c$ and referring to the temperatures employed in \cite{96ASGK}, we find that confinement effects are expected to appear when the diameter  satisfies $d \le 2.5$ nm. Thus the experiment had just missed the confinement effect by a hair. In Fig, \ref{PORE} we present our own
prediction as to how the expected position of $\omega_{max}$ depends on the confining pore diameter.
Note that there is no effect down to a critical pore diameter below which the shift in $\omega_{max}$  is dramatic. Here we are plotting shift as a function of pore diameter at fixed temperature. We also note
that as a function of temperature clusters tend to grow or shrink, and therefore confinement effects should be more dramatic at lower temperatures where the unconfined clusters are expected to be larger.

 \section{Discussion}
 \label{discussion}

We have shown how a relatively simple model of uncorrelated cluster 
relaxations can account, without much parameter-fitting, for the qualitative and  even the quantitative 
aspects of the observed BDS in hydrogen-bonded liquids. The theory was demonstrated for glycerol, but obviously with a mere change of molecular parameters it should apply to a broad range
of other hydrogen-bonded liquids. It was demonstrated explicitly that the $\alpha$ and $\beta$
regions of the dielectric spectra can in principle stem from the very same physics, and their relative amplitudes is determined by the relative population of small or large clusters. The relative population is determined by entropic effects which were explicitly taken into account in our thermodynamic theory, and by the surface energy per molecule in addition to the temperature. Applications to other materials will be presented elsewhere.
\begin{figure}
\centering
\epsfig{width=.5\textwidth,file=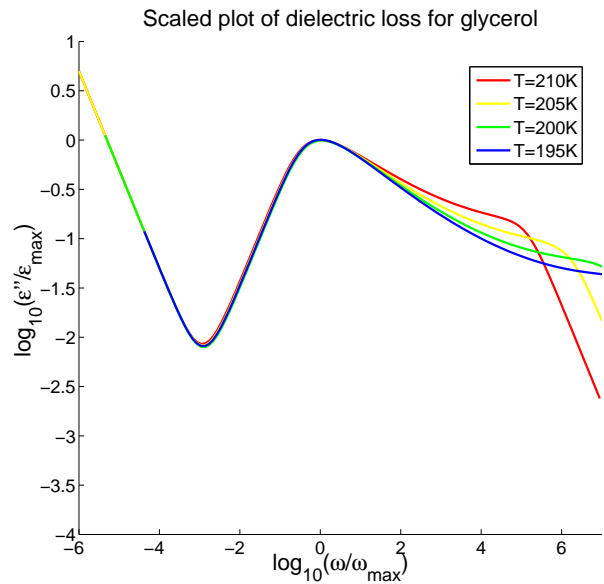}
\caption{Data collapse with the theoretical spectra: a rescaled version of the plots in Fig.~\ref{EPSILON} to which the dc conductivity contribution to the loss has been added.}
\label{SCALED}
\end{figure}

One particularly interesting aspect of the glycerol measurements, i.e. the data collapse shown in Fig.  \ref{rescaled}, was not used yet to challenge the theory. Note that the experimental data collapse includes the  clearly identified dc branch below frequencies $f/f_{max}\approx 10^{-3}$. 
We present our theoretical spectra in a similar manner, rescaling the frequency $\omega$ and the amplitude $\epsilon''(\omega)$ by the frequency and amplitude of the $\alpha$ peak. The result
of the exercise is shown in Fig. \ref{SCALED}, without any attempt to re-fit any of the material parameters. Note the approximate data collapse, with the excess wings failing to collapse perfectly.
It is clear that by choosing an appropriate value of $\sigma$ we could achieve a better data collapse, but
we do not try to do so here, deferring detailed quantitative fits to a future publication where theory
and experiments will be compared in full detail. We stress at this point that the dc branch was added here by hand, and we do not have a clear understanding why it is involved in the data collapse together
with the $\alpha$ peas and the `excess wing'.  The existence of such scaling at low frequencies suggests, however, that the dc conductivity can be written in the form $\sigma_{dc}\sim \epsilon_{max} \omega_{max}$. Why this is so requires further investigation.

\acknowledgements

This work had been supported in part by the German Israeli Foundation and the Minerva Foundation, Munich, Germany. The authors are grateful to Yuri Feldman and Alexander Puzenko for introducing
them to the subject and for sharing with them their knowledge and experimental results.

\end{document}